\documentclass[twocolumn,english,aps,prb,superscriptaddress,floatfix]{revtex4-2}
\usepackage[utf8]{inputenc}
\usepackage{xcolor}
\usepackage{babel}
\usepackage{bm}
\usepackage{amsmath}
\usepackage{amssymb}
\usepackage[bookmarks=false,
 breaklinks=true,pdfborder={0 0 1},backref=false,colorlinks=true]
 {hyperref}
\hypersetup{pdftitle={Fisher Geometry of Supercell Pair Distribution Function Refinement},
 pdfauthor={Dawei Wang},
 linkcolor=blue,citecolor=blue,urlcolor=blue}

\makeatletter
\usepackage{babel}

\usepackage{times}
\usepackage{bm}
\usepackage{booktabs}
\usepackage{graphicx}
\usepackage{xcolor}

\newcommand{\bG}{\bm{G}}
\newcommand{\bR}{\bm{R}}
\newcommand{\bd}{\bm{d}}

\newcommand{\bv}{\bm{v}}

\newcommand{\by}{\bm{y}}
\newcommand{\bDelta}{\bm{\Delta}}
\newcommand{\bSigma}{\bm{\Sigma}}
\newcommand{\bF}{\bm{F}}
\newcommand{\bJ}{\bm{J}}
\newcommand{\bA}{\bm{A}}

\newcommand{\bI}{\bm{I}}
\newcommand{\bB}{\bm{B}}
\newcommand{\bK}{\bm{K}}

\newcommand{\bH}{\bm{H}}

\newcommand{\KL}{D_{\mathrm{KL}}}
\newcommand{\E}{\mathbb{E}}

\newcommand{\dd}{\mathrm{d}}

\renewcommand{\top}{\intercal} 

\makeatother

\begin{document}
\title{A Fisher-Geometric Approach to Supercell Pair Distribution Function
Refinement}
\author{Dawei Wang}
\affiliation{Texas Center for Superconductivity and Department of Physics, University
of Houston, Houston, TX 77204, USA}
\begin{abstract}
Refining atomistic supercells against powder pair distribution functions
(PDFs) requires optimizing many coupled atomic coordinates using a
much smaller set of radial measurements. Reverse-mode differentiation
makes gradients accessible for large systems \cite{Yang2025Fitting},
but standard first-order optimizers do not explicitly use coupled
PDF sensitivities to precondition coordinate updates. We introduce
a matrix-free, Fisher-geometric approach that pulls the information
metric of the observation likelihood back onto atomic coordinates.
The resulting update weights collective displacements according to
their PDF response and experimental uncertainty. Under isotropic damping,
the update is obtained by solving a damped linear system over $M$
PDF channels, using automatic differentiation to evaluate Jacobian--vector
and vector--Jacobian products without constructing the full Jacobian
or coordinate curvature matrix. For Gaussian measurement errors, the
update recovers classical Levenberg--Marquardt refinement while avoiding
dense coordinate-space factorization. The same computational framework
accommodates correlated Gaussian errors and robust likelihood-based
updates. This approach combines likelihood-aware collective updates
with matrix-free backpropagation for direct second-order refinement
of large atomistic supercells. 
\end{abstract}
\maketitle
Refinement of atomic pair distribution functions (PDFs) from total
scattering measurements probes local structural distortions, chemical
short-range order, and nanoscale fluctuations in complex materials~\cite{EgamiBillinge2012,Keen2001}.
When structural disorder extends beyond an average unit cell, modeling
can require large periodic supercells containing $N\sim10^{4}\text{--}10^{6}$
atoms, making the treatment of $3N$ coordinate degrees of freedom
computationally demanding.

Zeroth-order trial-move techniques, exemplified by Reverse Monte Carlo
(RMC) \cite{McGreevy1988RMC}, refine large supercells by iteratively
testing coordinate displacements against diffraction data. However,
because experimental total scattering data provide an inherently underdetermined
constraint on extended Cartesian configurations, unconstrained stochastic
walks can explore structurally degenerate configurations and produce
unphysical coordination environments unless guided by prior constraints.
Classical least-squares refinement \cite{Proffen1999PDFFIT,Farrow2007PDFfit2}
incorporates interatomic sensitivities through the second-order normal
matrix under an assumed Gaussian likelihood (Levenberg--Marquardt).
However, assembling and factoring this dense $(3N)\times(3N)$ normal
matrix scales as $\mathcal{O}((3N)^{3})$ arithmetic and $\mathcal{O}((3N)^{2})$
memory for direct factorization, limiting the size of models accessible
to this dense formulation. Recently, first-order coordinate backpropagation
\cite{Yang2025Fitting} used reverse-mode automatic differentiation
to evaluate loss gradients for large supercells. However, standard
optimizers, such as Adam, adjust coordinate-wise step sizes using
gradient history without explicitly accounting for how displacements
of equal magnitude along different collective directions affect the
predicted PDF. While Euclidean distance measures how far the atoms
move, it does not quantify whether the resulting change in the predicted
PDF is experimentally detectable.

In total scattering, collective atomic displacements can produce different
changes in the diffraction pattern depending on interatomic vector
orientations, while extended displacement patterns may produce zero
diffraction response. Here, we derive the preconditioning from the
Fisher geometry of the observation likelihood. This formulation connects
Cartesian coordinate updates to experimental distinguishability and
avoids dense $(3N)\times(3N)$ coordinate inversion through a measurement-space
dual solve evaluated using forward- and reverse-mode automatic differentiation.

Consider a periodic supercell containing $N$ atoms at Cartesian coordinates
$\bR=(\bR_{1},\dots,\bR_{N})^{\top}\in\mathbb{R}^{3N}$, within a
periodic cell defined by the fixed lattice matrix $H$. At a synchrotron
or neutron beamline, experimental total scattering yields a reduced
pair distribution function sampled on a discrete radial grid of $M$
channels $\{r_{k}\}^{M}_{k=1}$ up to a real-space cutoff $r_{\max}$,
represented by the observation vector $\by\in\mathbb{R}^{M}$. The
forward model predicts the theoretical PDF $\bG(\bR)=(G_{1}(\bR),\dots,G_{M}(\bR))^{\top}\in\mathbb{R}^{M}$.
The observation likelihood $p(\by\mid\bR)\equiv p(\by\mid\bG(\bR))$
defines the distribution of noisy detector outcomes. The statistical
distinguishability between nearby predictions $\bG$ and $\bG+\delta\bG$
is quantified to quadratic order by the Kullback--Leibler divergence:
\begin{equation}
\begin{split}\dd s^{2} & =2\KL(p_{\bG}\,\|\,p_{\bG+\delta\bG})\\
 & \simeq\delta\bG^{\top}\bm{W}(\bG)\,\delta\bG=\delta\bR^{\top}\bF_{\bR}\,\delta\bR,
\end{split}
\label{eq:metric_unification}
\end{equation}
where $\bm{W}(\bG)\equiv\E_{\by}[-\nabla^{2}_{\bG}\ln p(\by\mid\bG)]$
is the Fisher information metric of the observation likelihood, defined
on the predicted PDF signal $\bG$. The Jacobian $\bJ_{G}\equiv\nabla_{\bR}\bG$
pushes forward the infinitesimal coordinate displacements $\delta\bR$
into $\delta\bG=\bJ_{G}\delta\bR$. Pulling the metric back onto coordinates
defines the coordinate Fisher information metric tensor $\bF_{\bR}\equiv\bJ^{\top}_{G}\bm{W}(\bG)\bJ_{G}=\E_{\by}[-\nabla^{2}_{\bR}\ln p(\by\mid\bR)]$.

For a given structure $\bR$, the distribution family $p(\by\mid\bR)$
determines the Fisher metric $\bF_{\bR}$ while the realized observation
$\by$ determines the loss and its optimization direction: 
\begin{equation}
\mathcal{L}_{\mathrm{obs}}(\bG)=-\ln p(\by\mid\bG).\label{eq:loss_potential}
\end{equation}
Minimizing $\mathcal{L}_{\mathrm{obs}}$ brings the predicted pattern
$\bG(\bR)$ toward the observations $\by$. Pulled back onto coordinates,
the unpreconditioned loss gradient is $\mathbf{g}\equiv\nabla_{\bR}\mathcal{L}_{\mathrm{obs}}=\bJ^{\top}_{G}\nabla_{\bG}\mathcal{L}_{\mathrm{obs}}$.
In large supercells, different atomic displacement modes change the
predicted PDF by different amounts. Small displacements along some
collective modes produce large PDF changes, whereas displacements
of equal magnitude along others produce negligible responses. This
anisotropy makes a single Euclidean step size difficult to tune.

To define a coordinate update $\delta\bR$, we construct a quadratic
surrogate model $\mathcal{Q}(\delta\bR)$ balancing first-order loss
reduction against Riemannian distance $\dd s^{2}$, regularized by
a physical reference metric $\bH_{0}$: 
\begin{equation}
\begin{split}\mathcal{Q}(\delta\bR) & =\mathcal{L}_{\mathrm{obs}}(\bG)+(\nabla_{\bR}\mathcal{L}_{\mathrm{obs}})^{\top}\delta\bR\\
 & \quad+\frac{1}{2}\delta\bR^{\top}(\bF_{\bR}+\lambda\bH_{0})\,\delta\bR,
\end{split}
\label{eq:proximal_surrogate}
\end{equation}
where $\bF_{\bR}$ weights steps by the distinguishability of the
resulting observation distributions $p(\by\mid\bG(\bR))$ and $p(\by\mid\bG(\bR+\delta\bR))$.
Here $\bH_{0}=\nabla^{2}_{\bR}U_{0}(\bR_{0})$ contains harmonic force
constants evaluated at an equilibrated reference configuration $\bR_{0}$
where $\nabla_{\bR}U_{0}\approx\mathbf{0}$. The reference potential
$U_{0}$ can, for example, be obtained from machine learning~\cite{Zhang2024PRL,Ouyang2023JPCC,Zhang2022PRB,Ouyang2023PRB,Ouyang2024PRMat,Ouyang2024APS}.
This term regularizes unobservable null-space modes where $\bF_{\bR}$
vanishes. Setting the gradient $\nabla_{\delta\bR}\mathcal{Q}=\mathbf{0}$
yields the regularized natural gradient update in coordinate space:
\begin{equation}
(\bF_{\bR}+\lambda\bH_{0})\,\delta\bR=-\bJ^{\top}_{G}\nabla_{\bG}\mathcal{L}_{\mathrm{obs}}.\label{eq:map_update}
\end{equation}
As a special case, under additive Gaussian detector noise with precision
$\bm{W}$, the loss specializes to weighted least-squares $\mathcal{L}_{\mathrm{obs}}=\frac{1}{2}\|\by-\bG(\bR)\|^{2}_{\bm{W}}$.
Linearizing the forward scattering model $\bG(\bR+\delta\bR)\approx\bG(\bR)+\bJ_{G}\delta\bR$
within this least-squares objective directly recovers $\mathcal{Q}(\delta\bR)$
as the classical damped Gauss--Newton surrogate, showing that classical
Levenberg--Marquardt~\cite{Levenberg1944,Marquardt1963} is a realization
of this general formulation (see below).

Direct inversion of the $(3N)\times(3N)$ matrix in Eq.~\eqref{eq:map_update}
would incur $\mathcal{O}((3N)^{3})$ cost. However, the measurement
dimension $M$ (i.e., the number of bins for the PDF curve) can be
orders of magnitude smaller than coordinate degrees of freedom ($M\sim10^{2}\text{--}10^{3}\ll3N\sim10^{4}\text{--}10^{6}$).
Under isotropic damping $\bH_{0}=\bI_{3N}$, we define the precision-weighted
Jacobian $\bA=\bm{W}^{1/2}\bJ_{G}\in\mathbb{R}^{M\times3N}$, where
$\bm{W}^{1/2}$ is the symmetric matrix square root. Then $\bF_{\bR}=\bA^{\top}\bA$.
Applying the Woodbury matrix push-through identity, $(\bA^{\top}\bA+\lambda\bI_{3N})^{-1}\bA^{\top}=\bA^{\top}(\bA\bA^{\top}+\lambda\bI_{M})^{-1}$,
recasts the $(3N)$-dimensional coordinate problem in Eq.~\eqref{eq:map_update}
into an $M$-dimensional measurement-space solve: 
\begin{align}
(\bA\bA^{\top}+\lambda\bI_{M})\,\bm{\alpha} & =-\bm{W}^{-1/2}\nabla_{\bG}\mathcal{L}_{\mathrm{obs}},\label{eq:dual_solve}\\
\delta\bR_{\mathrm{geom}} & =\bJ^{\top}_{G}(\bm{W}^{1/2}\bm{\alpha})\label{eq:seed_swap_update}\\
 & \equiv\operatorname{VJP}(\bJ_{G},\bm{W}^{1/2}\bm{\alpha}),\nonumber 
\end{align}
where $\bA\bA^{\top}=\bm{W}^{1/2}\bJ_{G}\bJ^{\top}_{G}\bm{W}^{1/2}$
is an $M\times M$ symmetric positive-semidefinite operator, rendered
positive-definite by the damping parameter $\lambda>0$.

With the raw observation score $\mathbf{s}_{\mathrm{grad}}\equiv-\nabla_{\bG}\mathcal{L}_{\mathrm{obs}}$,
the raw gradient-descent update is $\delta\bR_{\mathrm{grad}}=\eta\operatorname{VJP}(\bJ_{G},\mathbf{s}_{\mathrm{grad}})$,
whereas the geometric update Eq.~\eqref{eq:seed_swap_update} uses
the curvature-calibrated vector $\mathbf{s}_{\mathrm{geom}}\equiv\bm{W}^{1/2}\bm{\alpha}$.
Both schemes evaluate coordinate-space displacements via a reverse-mode
vector-Jacobian product (VJP)~\cite{Bradbury2018JAX}, whereas optimizers
such as Adam rescale individual gradient components using their past
statistics, without explicitly representing off-diagonal curvature.
The primary distinction lies in the adjoint vector: raw backpropagation
injects the uncalibrated score $\mathbf{s}_{\mathrm{grad}}$, whereas
the geometric update solves the compact $M$-dimensional dual system~\eqref{eq:dual_solve}
to inject the curvature-calibrated $\mathbf{s}_{\mathrm{geom}}$.

Equations~\eqref{eq:dual_solve}--\eqref{eq:seed_swap_update} require
solving the $M$-dimensional equation $(\bA\bA^{\top}+\lambda\bI_{M})\bm{\alpha}=\mathbf{b}$
for $\bm{\alpha}$. The system can be solved iteratively without explicitly
forming $\bA$ or $\bA\bA^{\top}$. Each Krylov iteration requires
only the action of $\bm{v}\mapsto(\bA\bA^{\top}+\lambda\bI_{M})\bm{v}$.
With $\bA=\bm{W}^{1/2}\bJ_{G}$, this action is evaluated matrix-free
by one reverse-mode contraction through $\bJ_{G}$, followed by one
forward-mode contraction: 
\begin{equation}
(\bA\bA^{\top})\bv=\bm{W}^{1/2}\operatorname{JVP}\Big(\bJ_{G},\operatorname{VJP}(\bJ_{G},\bm{W}^{1/2}\bv)\Big).\label{eq:scattering_echo}
\end{equation}

We next examine how atomic displacements change the PDF. In an atomistic
supercell, the forward scattering model evaluates interatomic distances
across the $M$ discrete radial channels $\{r_{k}\}^{M}_{k=1}$ up
to a real-space cutoff $r_{\max}$ \cite{Yang2025Fitting}: 
\begin{equation}
G_{k}(\bR)=-4\pi\rho_{0}r_{k}+\sum_{p}h_{kp}(d_{p}),\label{eq:pair_model}
\end{equation}
where $\rho_{0}=N/V$ is the average atomic number density, and the
sum runs over all ordered distinct atom pairs $p=(i,j)$ with $i\ne j$,
pair distance $d_{p}=\|\bR_{i}-\bR_{j}\|$, and unit bond vector $\widehat{\bd}_{p}=(\bR_{i}-\bR_{j})/d_{p}$.
The broadened peak profile $h_{kp}(d_{p})=(w_{p}/Nr_{k})K_{\sigma}(r_{k}-d_{p})$,
where the coherent scattering weight $w_{p}=b_{i}b_{j}/\langle b\rangle^{2}$,
incorporates Debye--Waller vibrations and instrumental resolution
through the Gaussian kernel: 
\begin{equation}
K_{\sigma}(x)\equiv\frac{1}{\sqrt{2\pi\sigma^{2}}}\exp\left(-\frac{x^{2}}{2\sigma^{2}}\right).\label{eq:kernel_def}
\end{equation}
Differentiating Eq.~\eqref{eq:pair_model} with respect to Cartesian
coordinates yields the linear response $\delta G_{k}$: 
\begin{equation}
\delta G_{k}=\sum_{p}h'_{kp}(d_{p})\,\widehat{\bd}_{p}\cdot(\delta\bR_{i}-\delta\bR_{j}).\label{eq:jacobian_factorization}
\end{equation}

Comparing Eq.~\eqref{eq:jacobian_factorization} with $\delta\bG=\bJ_{G}\delta\bR$
reveals that the forward scattering Jacobian factors geometrically
as $\bJ_{G}=\bK\bB$, and $\bF_{\bR}=\bB^{\top}\bK^{\top}\bm{W}\bK\bB$,
where the pair operator $\bB:\mathbb{R}^{3N}\to\mathbb{R}^{N_{\mathrm{pairs}}}$
projects $3N$-dimensional Cartesian displacements onto scalar pair
distances with $(\bB\delta\bR)_{p}=\widehat{\bd}_{p}\cdot(\delta\bR_{i}-\delta\bR_{j})$,
while the radial projection kernel $\bK:\mathbb{R}^{N_{\mathrm{pairs}}}\to\mathbb{R}^{M}$
superposes individual pair stretches into PDF radial channels via
elements $K_{kp}=h'_{kp}(d_{p})$, such that $\delta\bG=\bK(\bB\delta\bR)$.
This factorization separates coordinate kinematics from observation
projection: atomic displacements $\delta\bR$ change interatomic distances
via $\bB$, and radial projection $\bK$ maps those changes into the
PDF.

Because $\bm{W}$ is positive-definite across detector channels, an
atomic displacement produces no first-order PDF response if and only
if $\bJ_{G}\bv=\mathbf{0}$. Equation~\eqref{eq:jacobian_factorization}
identifies two distinct mechanisms for such dark modes: (i) $\bB\bv=\mathbf{0}$,
comprising displacement modes that preserve all interatomic pair distances
to first order (which include global rigid translations, alongside
any infinitesimal network flexes), and (ii) $\bB\bv\neq\mathbf{0}$
with $\bK(\bB\bv)=\mathbf{0}$, involving collective bond distortions
whose opposing expansions and contractions cancel under the spherical
powder average.

To relate these dark modes to structural order, consider an order
parameter $\eta(\bR)$ (such as ferroelectric polarization $P_{z}$
or octahedral tilt angle $\theta_{\mathrm{tilt}}$ in perovskites),
the gradient $\nabla_{\bR}\eta$ identifies the displacement mode
that increases the order parameter most rapidly per unit Cartesian
displacement. However, if this mode produces no first-order PDF response
($\bJ_{G}\nabla_{\bR}\eta=\mathbf{0}$), then the update according
to the raw gradient $\mathbf{g}\equiv\nabla_{\bR}\mathcal{L}_{\mathrm{obs}}=\bJ^{\top}_{G}\nabla_{\bG}\mathcal{L}_{\mathrm{obs}}$
leaves the order parameter unchanged to first order: 
\begin{align}
\delta\eta & \approx(\nabla_{\bR}\eta)^{\top}\delta\bR\nonumber \\
 & =-(\nabla_{\bR}\eta)^{\top}\left(\varepsilon\bJ^{\top}_{G}\nabla_{\bG}\mathcal{L}_{\mathrm{obs}}\right)\nonumber \\
 & =-(\bJ_{G}\nabla_{\bR}\eta)^{\top}\left(\varepsilon\nabla_{\bG}\mathcal{L}_{\mathrm{obs}}\right)=0,
\end{align}
where $\varepsilon$ sets the step size. Therefore, both raw gradient
descent and the geometric update $\delta\bR_{\mathrm{geom}}$ are
blind to these dark modes to first order. If an initial supercell
configuration possesses high symmetry where $\nabla_{\bR}\eta\in\ker\bJ_{G}$,
raw gradient descent and the isotropically damped geometric update
cannot initiate this order through their first-order response.

To determine whether a candidate distortion pattern $\mathbf{v}$
(normalized by unit root-mean-square displacement $\|\mathbf{v}\|^{2}/N=1$)
is experimentally detectable at finite amplitude $a$, we consider
the displacement response $\Delta\bG(a,\mathbf{v})\equiv\bG(\bR+a\mathbf{v})-\bG(\bR)$.
Under a Gaussian observation model with fixed covariance $\bSigma$,
the statistical distinguishability measured using precision $\bm{W}=\bSigma^{-1}$
is: 
\begin{equation}
\begin{split}\mathcal{D}(a,\mathbf{v}) & \equiv\left(\Delta\bG(a,\mathbf{v})^{\top}\bm{W}\,\Delta\bG(a,\mathbf{v})\right)^{1/2}\\
 & \approx|a|\left(\mathbf{v}^{\top}\bF_{\bR}\mathbf{v}\right)^{1/2}=|a|\|\operatorname{JVP}(\bJ_{G},\mathbf{v})\|_{\bm{W}},
\end{split}
\label{eq:snr_diagnostic}
\end{equation}
where the weighted norm $\|\mathbf{x}\|_{\bm{W}}\equiv(\mathbf{x}^{\top}\bm{W}\mathbf{x})^{1/2}$
evaluates to $(\sum_{k}|(\bJ_{G}\mathbf{v})_{k}|^{2}/\sigma^{2}_{k})^{1/2}$
for independent channel uncertainties $\sigma_{k}$, computed via
a directional derivative $\bJ_{G}\mathbf{v}\equiv\operatorname{JVP}(\bJ_{G},\mathbf{v})$.
When $\mathcal{D}(a,\mathbf{v})\ll1$, the scattering response lies
below experimental uncertainty, rendering such distortion difficult
to distinguish from the unperturbed structure at that amplitude. If
the first-order response vanishes ($\bJ_{G}\mathbf{v}=\mathbf{0}$),
distinguishability relies entirely on higher-order PDF changes, whose
leading contribution can be quadratic ($\propto a^{2}$). In practice,
such higher-order signatures may be confounded with thermal Debye--Waller
broadening or instrumental resolution unless these contributions are
independently constrained.

A mechanical reference metric provides a complementary measure of
displacement cost. The two curvature terms encode distinct physical
information: the pulled-back Fisher metric $\bF_{\bR}=\bJ^{\top}_{G}\bm{W}\bJ_{G}$
quantifies experimental distinguishability, penalizing coordinate
motions according to how strongly they alter the measured PDF relative
to detector uncertainty, whereas a physical force-constant Hessian
$\bH_{0}=\nabla^{2}_{\bR}U_{0}(\bR_{0})$ from an equilibrated reference
potential supplies mechanical stiffness along $\ker\bJ_{G}$, penalizing
unobservable distortions according to bonding stiffness without exerting
baseline forces. When measurement curvature dominates the regularization,
the update approaches the unregularized Gaussian Gauss--Newton step;
along weakly observed directions, the regularizing metric suppresses
noise-induced step divergence, while along exact null directions the
mechanical metric supplies restoring stiffness, and translational
gauge degrees of freedom are eliminated by pinning the center of mass.
In this way, the measurement metric governs what the diffraction data
can resolve, while the mechanical reference metric regularizes directions
where the data are blind.

Specifying the observation likelihood $p(\by\mid\bG(\bR))$ determines
the weighting matrix $\bm{W}$ and score vector $\mathbf{s}$. Under
independent Gaussian noise with variance $\sigma^{2}_{k}$, the likelihood
is: 
\begin{equation}
p(\by\mid\bG(\bR))=\prod^{M}_{k=1}\frac{1}{\sqrt{2\pi\sigma^{2}_{k}}}\exp\left(-\frac{(y_{k}-G_{k}(\bR))^{2}}{2\sigma^{2}_{k}}\right).\label{eq:gauss_likelihood}
\end{equation}
The loss becomes weighted least-squares $\mathcal{L}_{\mathrm{Gauss}}=\frac{1}{2}\sum_{k}\Delta^{2}_{k}/\sigma^{2}_{k}$
with residual $\Delta_{k}=y_{k}-G_{k}$, and its gradient is $-\nabla_{\bG}\mathcal{L}_{\mathrm{Gauss}}=\bm{W}_{\mathrm{Gauss}}\bDelta$,
where the precision matrix is diagonal: $\bm{W}_{\mathrm{Gauss}}=\operatorname{diag}(\sigma^{-2}_{1},\dots,\sigma^{-2}_{M})$.
In this case, setting $\bA=\bm{W}^{1/2}_{\mathrm{Gauss}}\bJ_{G}$,
the dual solve specializes to: 
\begin{align}
(\bA\bA^{\top}+\lambda\bI_{M})\bm{\alpha} & =\bm{W}^{1/2}_{\mathrm{Gauss}}(\by-\bG(\bR)),\label{eq:dual_lm}\\
\delta\bR_{\mathrm{LM}} & =\operatorname{VJP}(\bJ_{G},\bm{W}^{1/2}_{\mathrm{Gauss}}\bm{\alpha}).\nonumber 
\end{align}
This recovers the classical Levenberg--Marquardt update \cite{Proffen1999PDFFIT,Farrow2007PDFfit2}
in measurement space. The update can be evaluated either by a direct
$M\times M$ Cholesky solve or by matrix-free Krylov iterations, followed
by a reverse-mode backpropagation sweep to compute the atomic displacements.

Beyond independent counting statistics, errors between radial bins
in experimental PDF analysis are correlated due to reciprocal-to-real
space Fourier data processing. The experimental PDF is computed from
reduced structure function data $F(Q)=Q[S(Q)-1]$ via a sine Fourier
integral up to $Q_{\max}$~\cite{EgamiBillinge2012,Qiu2004PDFgetX2}:
\begin{equation}
G(r)=\frac{2}{\pi}\int^{Q_{\max}}_{0}F(Q)\sin(Qr)\,\dd Q.\label{eq:fourier_integral}
\end{equation}
A fluctuation in reciprocal-space data can affect several radial channels
after Fourier transformation. These channel deviations should therefore
not be treated as independent evidence of a structural mismatch. Letting
$\bm{T}$ denote the linear discrete data transformation mapping $Q$-space
observations to real-space channels, the real-space error covariance
transforms as: 
\begin{equation}
\bSigma_{r}=\bm{T}\bSigma_{Q}\bm{T}^{\top}.\label{eq:general_cov_propagation}
\end{equation}
Under independent reciprocal counting statistics ($\bSigma_{Q}\approx\sigma^{2}_{Q}\bI$),
integrating Eq.~\eqref{eq:fourier_integral} yields characteristic
sinc-wave correlations with correlation length $\xi_{r}\sim\pi/Q_{\max}$.
Because adjacent radial bins are strongly correlated, treating them
as independent measurements inflates confidence and distorts parameter
uncertainties. While crystallographic refinements traditionally adopt
diagonal weighting for simplicity, our formulation in Eqs.~\eqref{eq:dual_solve}--\eqref{eq:seed_swap_update}
naturally accommodates the full off-diagonal covariance. Correlated
errors can be incorporated through $\bm{W}=\bSigma^{-1}_{r}$, with
an $M\times M$ Cholesky factorization of $\bSigma_{r}$ providing
the required whitening operations directly within the measurement-space
solve, without forming additional coordinate-space matrices.

Beyond correlated Gaussian noise, measured scattering data can exhibit
outliers or deviations from Gaussian errors. Under a Gaussian likelihood,
the least-squares observation score scales linearly with the residual
($s^{\mathrm{Gauss}}_{k}=\Delta_{k}/\sigma^{2}_{0}$), allowing large
outliers to disproportionately influence the structural refinement.
If channel reliability is uncertain, an anomalously large residual
should not automatically exert an unbounded pull on atomic coordinates.
To accommodate outliers or deviations from Gaussian errors, we consider
a Student-$t$ observation model, which accommodates heavy-tailed
residuals through degrees of freedom $\nu$ and scale parameter $\sigma_{0}$.
Small values, such as $\nu=2\text{--}4$, produce heavy tails, while
$\sigma_{0}$ matches baseline measurement uncertainty: 
\begin{equation}
\mathcal{L}_{t}(\bR)=\frac{\nu+1}{2}\sum^{M}_{k=1}\ln\left[1+\frac{\Delta_{k}(\bR)^{2}}{\nu\sigma^{2}_{0}}\right],\label{eq:student_loss}
\end{equation}
which produces an observation score vector $\mathbf{s}_{t}\equiv-\nabla_{\bG}\mathcal{L}_{t}$
with dynamic weights: 
\begin{equation}
(s_{t})_{k}=(W_{t})_{kk}\Delta_{k},\qquad(W_{t})_{kk}=\frac{\nu+1}{\nu\sigma^{2}_{0}+\Delta^{2}_{k}}.\label{eq:student_weights}
\end{equation}
For small residuals ($|\Delta_{k}|\ll\sigma_{0}$), $(W_{t})_{kk}\approx\frac{\nu+1}{\nu\sigma^{2}_{0}}\sim\sigma^{-2}_{0}$,
recovering a locally Gaussian-like linear response ($s_{t}\propto\Delta_{k}$).
Conversely, for large deviations ($|\Delta_{k}|\gg\sigma_{0}$), the
denominator is dominated by $\Delta^{2}_{k}$, so that the score tends
to zero as the residual magnitude increases, thereby reducing the
influence of large residuals on the coordinate update $\bJ^{\top}_{G}\bm{s}_{t}$.

For the Student-$t$ likelihood with fixed $\nu$ and $\sigma_{0}$,
the expected Fisher information is $\bm{W}_{\mathrm{Fisher}}=\frac{\nu+1}{(\nu+3)\sigma^{2}_{0}}\bI_{M}$~\cite{Hartmann2019},
which does not depend on realized residuals. In robust refinement,
we instead use the residual-dependent weights $\bm{W}_{t}$ as a positive
iteratively reweighted surrogate curvature. Freezing $\bm{W}_{t}$
at each iteration within Eqs.~\eqref{eq:dual_solve}--\eqref{eq:seed_swap_update}
yields the update: 
\begin{align}
(\bA_{t}\bA^{\top}_{t}+\lambda\bI_{M})\bm{\alpha} & =\bm{W}^{1/2}_{t}\bDelta,\label{eq:dual_rlm}\\
\delta\bR_{\mathrm{RLM}} & =\operatorname{VJP}(\bJ_{G},\bm{W}^{1/2}_{t}\bm{\alpha}),\nonumber 
\end{align}
where $\bA_{t}=\bm{W}^{1/2}_{t}\bJ_{G}$. This scheme implements iteratively
reweighted Levenberg--Marquardt refinement \cite{Triggs2000,Lange1989Robust}
in measurement space. Evaluating the update in measurement space requires
an $M\times M$ linear solve followed by a reverse-mode backpropagation
sweep, sharing the same leading per-iteration computational structure
as the Gaussian case.

These three observation models demonstrate a unified principle: regardless
of the specific likelihood or experimental error assumptions, the
structural update adheres to the identical measurement-space dual
architecture in Eqs.~\eqref{eq:dual_solve}--\eqref{eq:seed_swap_update}
through the choice of metric or surrogate weighting $\bm{W}$. Recovering
classical and robust Levenberg--Marquardt updates confirms consistency
with established least-squares methods, while bypassing the formation
and storage of dense coordinate-space normal matrices. Evaluating
updates through matrix-free operations and the measurement-space dual
solve makes second-order preconditioning computationally practical
for supercell PDF modeling, complementing stochastic trial-move approaches~\cite{McGreevy1988RMC}
and first-order backpropagation~\cite{Yang2025Fitting}.

This construction shares the natural-gradient principle underlying
stochastic reconfiguration (SR) in quantum variational Monte Carlo~\cite{Sorella1998Green,Sorella2005Wave,Amari1998Natural,Martens2020New}:
parameter updates are scaled by a metric describing distinguishability
in the represented state or observation space. In both settings, the
loss provides the driving force, while the metric governs the step
geometry. The mathematical objects differ, however: quantum SR employs
the Fubini--Study metric of variational wavefunctions sampled from
Monte Carlo configurations, whereas the present approach combines
the Fisher metric of observation distributions $p(\by\mid\bR)$ with
deterministic scattering derivatives evaluated by automatic differentiation.
Optimizing an ensemble distribution $q_{\phi}(\bR)$ over atomic configurations
would introduce a separate geometry in configuration-distribution
space.

The same likelihood-geometric approach extends naturally to single-crystal
3D-$\Delta\mathrm{PDF}$ and three-dimensional diffuse-scattering
refinement~\cite{Weber2012,Juul2025} through an appropriate differentiable
forward model. In single-crystal diffuse scattering, measuring full
three-dimensional reciprocal volumes directly captures directional
interatomic correlations, which can distinguish directional correlations
lost under powder averaging. In localized difference volumes and region-of-interest
reconstructions where measurement channels remain smaller than coordinate
degrees of freedom ($M_{\mathrm{ROI}}\ll3N$), the measurement-space
dual solve retains its dimensional advantage.

In summary, pulling back the Fisher information metric of the observation
likelihood onto atomic coordinates anchors collective structural updates
to experimental distinguishability. This formulation separates observable
structural distortions from powder-blind dark modes, allowing a mechanical
reference metric to regularize unconstrained directions rather than
relying solely on isotropic damping. Automatic differentiation through
JVPs and VJPs provides a practical, matrix-free route to collective
coordinate updates in supercell PDF refinement. Combined with the
measurement-space dual solve, this approach avoids the coordinate-space
scaling of classical Levenberg--Marquardt, providing a practical
route to second-order refinement for large supercells across powder
diffraction and three-dimensional diffuse scattering. 
\begin{acknowledgments}
This work was supported by the Texas Center for Superconductivity
at the University of Houston (TcSUH) and The Robert A. Welch Foundation
under Grant No. L-E-0001-19921203. The author also acknowledges the
computing resources provided by the HPE Data Science Institute and
the Center for Advanced Computing and Data Systems at the University
of Houston. 
\end{acknowledgments}

\bibliography{references}

\end{document}